\documentclass[aps,prb,twocolumn,amsmath,showpacs,amssymb,superscriptaddress]{revtex4}
\usepackage{amsmath}
\usepackage{graphicx}
\usepackage{amssymb}
\usepackage{dsfont}

\newcommand{\pdag}{{\phantom{\dagger}}}

\begin{document}

\title{Low-energy properties of fractional helical Luttinger liquids}

\author{Tobias Meng}
\affiliation{Department of Physics, University of Basel, Klingelbergstrasse 82, CH-4056 Basel, Switzerland}

\author{Lars Fritz}
\affiliation{Institut f\"ur Theoretische Physik, Universit\"at zu K\"oln, Z\"ulpicher Stra\ss e 77, 50937 K\"oln, Germany}

\author{Dirk Schuricht}
\affiliation{Institut f\"{u}r Theorie der Statistischen Physik, RWTH Aachen University and JARA - Fundamentals of Future Information Technology, 52056 Aachen, Germany}

\author{Daniel Loss}
\affiliation{Department of Physics, University of Basel, Klingelbergstrasse 82, CH-4056 Basel, Switzerland}

\begin{abstract}
We investigate the low-energy properties of (quasi) helical and fractional helical Luttinger liquids. In particular, we calculate the Drude peak of the optical conductivity, the density of states, as well as charge transport properties of the interacting system with and without attached Fermi liquid leads at small and large (compared to the gap) frequencies. For fractional wires, we find that the low energy tunneling density of states vanishes. The conductance of a fractional helical Luttinger liquid is non-integer. It is independent of the Luttinger parameters in the wire, despite the intricate mixing of charge and spin degrees of freedom, and only depends on the relative locking of charge and spin degrees of freedom.
\end{abstract}

\maketitle

\section{Introduction}
Being of great use in the context of Majorana bound states in topological quantum wires,\cite{majorana_review} and having applications such as Cooper pair splitters\cite{cooper_splitter} or spin filters,\cite{streda_03} helical and quasi helical quantum wires have been intensely studied both theoretically and experimentally in recent years. They have been proposed to emerge as edge channels of quantum spin Hall systems,\cite{kane_mele} in quantum wires made from three-dimensional topological insulator materials,\cite{ti_wire1,ti_wire2,ti_wire3} in quantum wires with helical nuclear spin order,\cite{braunecker_prb_09,braunecker_prl_09, meng_13, scheller_12} in Rashba nanowires subject to a magnetic field when the chemical potential exactly coincides with the energy of the lifted band crossing at zero momentum,\cite{streda_03,charis_gaas_soi_gap_10}  in nanowires in a spatially oscillating (rotating) magnetic field, \cite{braunecker_prb_10,klinovaja_12} or in Carbon nanotubes.\cite{klinovaja_helical_cnt_11} In a recent work, a similar partially gapped Luttinger liquid model has been proposed as a realization of a fractional helical Luttinger liquid.\cite{oreg_fractional_13} This state can be understood as the one-dimensional analogue of a model proposed in Ref.~[\onlinecite{kane_02}], which described the fractional quantum Hall effect through a collection of coupled, interacting quantum wires. In this sense, fractional helical Luttinger liquids are related to the edge states of fractional quantum Hall systems.\cite{oreg_fractional_13}

One experimental signature of helical Luttinger liquids is provided by its charge transport properties.\cite{scheller_12,charis_gaas_soi_gap_10} While a reduced conductance as compared to a regular spinful Luttinger liquid is certainly characteristic for a helical state, it does not unambiguously identify this state. A number of recent works has thus addressed further properties of a helical Luttinger liquid with an emphasis on its spectral properties\cite{bena_12,schuricht_12}, disorder effects,\cite{braunecker_disorder} and its spin response.\cite{meng_spin_13} In this work, we further investigate the spectral properties of (quasi) helical and fractional helical Luttinger liquids, thereby also taking into account the coupling between gapped and gapless modes in the wire, and address its response at small but finite frequencies. By calculating various low-energy characteristics, we find that the charge transport properties of a (quasi) helical or a fractional helical Luttinger liquid can be understood as the ones of a regular spinless Luttinger liquid coupled to (interacting) Luttinger liquid leads. Their conductance is independent of the Luttinger parameters in the wire, but depends on the form of the sine-Gordon potential that opens the partial gap, as has already been found in Ref.~[\onlinecite{oreg_fractional_13}]. The analogy between a partially gapped and a spinless Luttinger liquid does, however, not hold for the electronic tunneling density of states, which vanishes for fractional helical Luttinger liquids.

We organize this manuscript as follows. In Sec.~\ref{sec:model}, we introduce the general model describing fractional helical Luttinger liquids. Sec.~\ref{sec:inifite_wire} is devoted to the analysis of the low-energy conductance, conductivity, and density of states of an infinite helical Luttinger liquid, while the conductance of a finite size wire attached to Fermi liquid leads is addressed in Sec.~\ref{sec:FLL}. In Sec.~\ref{sec:finite_freq}, we finally investigate the finite frequency conductivity of a fractional helical Luttinger liquid and discuss the asymptotic regimes of the frequencies being much smaller or much larger than the gap.

\section{The model}\label{sec:model}
We consider a one-dimensional electron system in which spin and charge excitations are locked by a sine-Gordon potential. One currently well-studied example are Rashba nanowires in a magnetic field. The Rashba interaction defines a spin quantization axis, and splits the dispersions for spin up and spin down in momentum space. A homogenous magnetic field applied perpendicular to the Rashba-defined spin quantization axis then allows for spin flips and opens up a partial gap close to zero momentum.\cite{streda_03} The residual gapless modes are helical in that their spin is (approximately) aligned to their direction of motion. Denoting the spin quantization axis set by the Rashba interaction as $\hat{z}$ in spin space, the perpendicular magnetic field can be chosen to point along the spin-$\hat{x}$ axis. Its Hamiltonian thus takes the form

\begin{align}
H_{\rm B} &=\int dx\,\boldsymbol{S}(x)\cdot\boldsymbol{B}=\int dx\,\frac{c_{\uparrow}^\dagger(x)\,c_{\downarrow}^\pdag(x)}{2}\,B+\text{h.c.}~,
\end{align}
where $\boldsymbol{S}(x)$ is the electron spin at position $x$ along the wire, $\boldsymbol{B}=B\,\hat{x}$ denotes the applied magnetic field, while $c_\nu^\pdag(x)$ annihilates an electron of spin $\nu=\uparrow,\downarrow$ at position $x$. In order to distill a bosonized version of this Hamiltonian, the magnetic field is treated as a perturbation to the ungapped Rashba nanowire. We linearize the latter's kinetic energy around the Fermi points and decompose the fermionic operators into right and left movers as $c_{\nu}(x) = e^{i x(k_F+\nu\, k_{SO})} R_{\nu}(x) + e^{-i x(k_F-\nu \,k_{SO})} L_{\nu}(x)$ with $\nu=\uparrow,\downarrow\equiv\pm1$, where $k_{SO}$ is the spin-orbit momentum, while $k_F$ corresponds to the Fermi momentum in the absence of spin-orbit interaction (we assume $k_{SO}>0$ without loss of generality). Bosonizing the theory as usual,\cite{giamarchi_book} the magnetic field translates to a sine-Gordon term. For a general value of the chemical potential, this sine-Gordon term contains rapidly oscillating exponential factors $\sim \text{exp}(2ix(k_F\pm k_{SO}))$, which render the magnetic field irrelevant. Only when the chemical potential is placed in the vicinity of the band crossing at zero momentum, that is for $k_F\approx k_{SO}$, the sine-Gordon term has the non-oscillating part
\begin{align}
H_{\rm B} &=\int dx\, \frac{B}{2\pi a}\,\cos(\sqrt{2}(\phi_c(x)+\theta_s(x)))~,\label{eq:rashba_ham}
\end{align}
where $\phi_c(x)$ is the bosonic field related to the integrated charge density, while $\theta_s(x)$ is proportional to the integrated spin current, and $a$ denotes a short-distance cutoff. This sine-Gordon term is in general relevant in the renormalization group (RG) sense, and its flow to strong coupling signals the opening of the before-mentioned partial gap in the bosonic language.

The mixing of the bosonic spin and charge fields may, however, take more complicated forms than the one in Eq.~\eqref{eq:rashba_ham}. As such, it has been proposed\cite{oreg_fractional_13,kane_02} that higher order interaction terms can give rise to sine-Gordon potentials that generalize the spin-charge mixing of Rashba nanowires, and lead to the emergence of a fractional helical Luttinger liquid with fractional electric conductance (in this denomination, a Rashba nanowire corresponds to an integer helical Luttinger liquid). If brought into contact with a superconductor, these systems are expected to host fractionalized Majorana bound states (parafermions).\cite{oreg_fractional_13} A fractional helical Luttinger liquids can be constructed by combining the $(2n)^{\rm th}$ order of a perturbation theory in a (contact) interaction $U$ with the magnetic field, which generates a term of the form

\begin{align}
H_{{\rm B},n} &\sim B\,U^{2n}\,\int dx\,\biggl(e^{-i2(k_F-k_{SO})}\,L_{\uparrow}^\dagger(x)\,R_{\downarrow}^\pdag(x)\label{eq:sine-gordon_gen}\\
&\times\,e^{-i4nk_F} \left(R_{\uparrow}^\dagger(x)L_{\uparrow}^\pdag(x)R_{\downarrow}^\dagger(x)L_{\downarrow}(x)\right)^n\biggr)+\text{h.c.}~,\nonumber
\end{align}
This term is non-oscillating if the chemical potential is fine-tuned such that $(2n+1)\,k_F = k_{SO}$, in which case the bosonization of Eq.~\eqref{eq:sine-gordon_gen} yields $H_{{\rm B},n}\sim\int dx\cos(\sqrt{2}([2n+1]\phi_c+\theta_s))$.\cite{oreg_fractional_13} The non-conservation of spin furthermore allows for an interaction of the type $R_{\downarrow}^\dagger L_{\uparrow}^\pdag L_{\downarrow}^\dagger R_{\uparrow} \sim \text{exp}(-i4k_{SO})\,\text{exp}(i2\sqrt{2}\theta_s)$. Combining this interaction with the former one, it is possible to generate sine-Gordon terms of the form $\cos(\sqrt{2}([2n+1]\phi_c+[2m+1]\theta_s))$ with $n,m\in\mathds{Z}$, provided the chemical potential is fine-tuned to $(1+2n)\,k_F = (1+2m)\, k_{SO}$. 

As an alternative route to the generation of a fractional helical Luttinger liquid, one may also envision the quasi one-dimensional analogue of the construction used in Ref.~[\onlinecite{kane_02}], namely two coupled quantum wires of spinless electrons. If a magnetic field is applied perpendicular to the plane of the wires, each tunneling events between the wires is associated with a momentum kick. If this kick matches the momentum difference between right movers of the first wire and left movers of the second wire, a partial gap analogue to the helical gap in a Rashba nanowire opens.\cite{kane_02,braunecker_prb_10} A combined perturbation theory in the Coulomb repulsion and the pair tunneling between the wires then again allows to promote the inter wire tunneling to the form $\sim\int dx\cos(\sqrt{2}([2n+1]\phi_c+[2m+1]\theta_s))$, where the spin field $\theta_s$ is now defined with respect to the pseudospin labeling the two wires. Again, a given sine-Gordon term is only stable for a specific value of $k_F$ at fixed perpendicular magnetic field. These values can be understood as the one-dimensional analogues of the filling fractions of the fractional quantum Hall state.\cite{kane_02,oreg_fractional_13}

In this work, we aim at analyzing the low-energy properties of generic fractional helical Luttinger liquids. To this end, we consider a general model with the imaginary time action (in units of $\hbar = 1$)\cite{braunecker_prb_10,braunecker_prb_09,braunecker_prl_09, meng_13,oreg_fractional_13}
\begin{align}
\mathcal{S}&=\frac{1}{2}\int dx \, d\tau \phi_c (x,\tau) \,G^{-1}_{cc}(x,\tau) \,\phi_c (x,\tau) \nonumber \\ &+ \frac{1}{2}\int dx \, d\tau \, \theta_s (x,\tau) \,G^{-1}_{ss}(x,\tau)\,\theta_s(x,\tau)  \label{eq:general_action}\\ &+  \int dx \, d\tau\, \frac{B(x)}{2\pi a}\cos \left( \sqrt{2}\left( \gamma_c \phi_c(x,\tau) + \gamma_s \theta_s(x,\tau)\right) \right)\nonumber~.
\end{align}
Here, we have introduced the inverse propagators $G^{-1}_{cc}=-\partial_x \frac{v_c(x)}{\pi K_c(x)} \partial_x - \frac{1}{ \pi v_c(x) K_c(x)}\partial^2_\tau$ and $G^{-1}_{ss}=-\partial_x \frac{v_s(x) K_s(x)}{\pi} \partial_x - \frac{K_s(x)}{\pi v_s(x)}\partial^2_\tau$. As before, the bosonic fields $\phi_c$ and $\theta_s$ relate to the integrated charge density and spin current, respectively, $v_c$ and $v_s$ are the effective velocities of the charge and spin excitations, while $K_c$ and $K_s$ are the corresponding Luttinger parameters. The locking of charge and spin degrees of freedom is described by the parameters $\gamma_c$ and $\gamma_s$. The relative locking $\delta=\gamma_s/\gamma_c$ may also be understood as a measure of the degree of fractionalization in the wire,\cite{oreg_fractional_13} see below. If the argument of the cosine potential derives from a combination of an integer number interaction terms and the backscattering term, $\gamma_c$ and $\gamma_s$ can take odd integer values, see above. Rashba nanowires and, more generally, spiral Luttinger liquids\cite{braunecker_prb_09,braunecker_prl_09,bena_12,braunecker_disorder,meng_13,schuricht_12} have $\gamma_c, \gamma_s = \pm1$, while the fractional Luttinger liquids considered in Ref.~[\onlinecite{oreg_fractional_13}] have $\gamma_s = 1$ and  $\gamma_c = (2n+1)$. All parameters can exhibit a spatial dependence, which allows to model the coupling of the interacting wire to Fermi liquid leads, see Sec.~\ref{sec:FLL}.

The sine-Gordon term proportional to $B$ is most conveniently analyzed by a renormalization group (RG) analysis and may at low energies either scale to zero or open up a gap. Since the fixed point action in the former case corresponds to a gapless and therefore trivial quantum wire, we assume the sine-Gordon potential to be relevant in the RG sense. An analysis along the lines of references [\onlinecite{braunecker_prb_09,braunecker_prl_09, meng_13}] shows that a gap is indeed opened for $K_c<(4-\gamma_s^2/K_s)/\gamma_c^2$, while the fixed point action is gapless for larger values of $K_c$. Electron-electron interactions are thus not only crucial in generating the sine-Gordon terms, but also decide on their relevance in the RG sense. Once the gap is opened, the wire's low-energy properties can be described by expanding the cosine potential to quadratic order in the fields. In the limit of a large gap, this quadratic expansion is equivalent to a self-consistent harmonic approximation. This yields the low-energy action
\begin{align}\label{eq:lowenergy_action}
\mathcal{S}_{\rm{eff}}=\frac{1}{2\pi}\int dx \,d\tau \,\boldsymbol{\Psi}^T \left( \begin{array}{cc}\hat{\mathcal{D}}_c(x,{\tau})& \delta \Delta^2(x)  \\ \delta \Delta^2(x)  & \hat{\mathcal{D}}_s(x,{\tau})\end{array}\right) \boldsymbol{\Psi}~,
\end{align}
where $\Psi = (\phi_c,\theta_s)^T$ and
\begin{subequations}
 \begin{align}
\hat{\mathcal{D}}_c(x,{\tau})&=-\partial_x\,\frac{v_c(x)}{K_c(x)}\,\partial_x-\frac{\partial_\tau^2}{ v_c(x) K_c(x)}+\Delta^2(x) ~,\\
  \hat{\mathcal{D}}_s(x,{\tau})&=-\partial_x\,v_s(x)K_s(x)\,\partial_x-\frac{K_s(x)}{v_s(x)}\partial_\tau^2+\delta^2\Delta^2(x) ~,
\end{align}\label{eq:diff_op}
\end{subequations}
and with $\Delta^2(x)=\gamma_c^2B(x)/a$. This simplified action provides the starting point for the subsequent discussions. It is valid at energies much smaller than the gap, see Secs.~\ref{sec:inifite_wire}-\ref{subsec:smaller}, but can also capture the regime of energies much larger than the gap, where the sine-Gordon potential constitutes a subleading energy scale (see Sec.~\ref{subsec:larger}). We can, however, not hope to reliably address the regime of energies of the order of the gap, where the sine-Gordon term should not be expanded (see Sec.~\ref{subsec:inter}).

\section{Infinite wire conductance, Drude peak, and density of states exponent}\label{sec:inifite_wire}
In a first step, we investigate the conductance and the Drude peak of an infinite and homogenous wire. Both of these quantities follow from taking the limit $\omega \to 0$ on the  retarded charge propagator. For this limit, it is sufficient to Fourier transform Eq.~\eqref{eq:lowenergy_action} to momentum $q$ and Matsubara frequencies $\omega_n$, to then invert the resulting matrix, and to finally perform the limit $\Delta \to \infty$. Finite $\Delta$ results in corrections of the order $\omega_n/(\Delta \sqrt{v_c})$ and $\omega_n/(\Delta\sqrt{v_s})$, which vanish at zero frequency.

The infinite gap limit of the charge propagator reads
\begin{align}
G_{cc}(k,\omega_n)=\frac{\pi}{\left(\frac{v_c}{K_c}+\frac{v_s K_s}{\delta^2} \right) k^2 + \left (\frac{1}{v_c K_c}+ \frac{K_s}{v_s \delta^2} \right) \omega_n^2}~.\label{eq:gcc_delta}
\end{align}
In this limit, the spin propagator and the mixed propagators are furthermore identical up to signs and factors of $\delta$,
\begin{align}
G_{cc}=G_{ss}/\delta^2=-G_{cs}/\delta=-G_{sc}/\delta\;.
\end{align}
One can now rewrite the theory with an effective Luttinger parameter $K_{\rm{eff}}$ and an effective velocity $v_{\rm{eff}}$,
\begin{subequations}\label{eq:effparam}
\begin{align}\label{eq:Keff}
K_{\rm{eff}}&= \frac{K_c \sqrt{v_c v_s}\delta^2}{\sqrt{(K_c K_s v_c+v_s \delta^2)(v_c \delta^2+K_c K_s v_s)}}~,  \\ v_{\rm{eff}}&= \frac{(v_c \delta^2+v_s K_c K_s) \sqrt{v_c v_s}}{\sqrt{(K_c K_s v_c+v_s \delta^2)(v_c \delta^2+K_c K_s v_s)}}~.
\end{align}
\end{subequations}
These expressions have also been found in reference [\onlinecite{braunecker_prb_09}].\cite{prb_remark} The Luttinger parameters and velocities appearing in Eqs.~\eqref{eq:effparam} are to be understood as the renormalized values obtained at the end of the flow associated with the RG relevant sine-Gordon potential $\sim B(x)$ mixing spin and charge degrees of freedom. In terms of the effective parameters, the charge propagator takes the form $G_{cc}(k,\omega_n) =\pi K_{\rm eff} v_{\rm eff}/(v_{\rm eff}^2 k^2+\omega_n^2)$. The physics of the charge sector may therefore be understood as the one of a spinless Luttinger liquid upon redefining $\tilde{\phi} = \sqrt{2}\,\phi_c$. The field $\tilde{\phi}$ then has the propagator $G_{\tilde{\phi}\tilde{\phi}}(k,\omega_n) =\pi \widetilde{K}_{\rm eff} v_{\rm eff}/(v_{\rm eff}^2 k^2+\omega_n^2)$ with $\widetilde{K}_{\rm eff} = 2 K_{\rm eff}$. This implies that the individual spin and charge velocities and Luttinger parameters have no physical importance once spin and charge degrees of freedom get locked. Instead, the single gapless mode remaining after the opening of the partial gap is characterized by a velocity $v_{\rm eff}$ and a Luttinger parameter $K_{\rm eff}$ given in Eq.~\eqref{eq:effparam}, which can for instance be accessed through the Drude peak or the tunneling density of states, see below, as well as spectroscopy experiments.\cite{auslaender_spectroscopy_02} The initial spin and charge parameters become visible only at energies much larger than the gap, at which spin and charge degrees of freedom can again be excited individually.

\subsubsection{Infinite wire conductance}
The conductance follows from the spatial Fourier transform of the analytic continuation of Eq.~\eqref{eq:gcc_delta}. In linear response theory, it is given by the Kubo formula

\begin{align}
 G &= \left.\frac{2e^2}{\pi^2}\,\omega_n\, \, G_{cc}(x,\omega_n)\right|_{i\omega_n \to \omega+i0^+,~\omega\to0}\label{eq:cond}\\
 &=\left.\frac{e^2}{\pi^2}\,\omega_n\, \, G_{\tilde{\phi}\tilde{\phi}}(x,\omega_n)\right|_{i\omega_n \to \omega+i0^+,~\omega\to0} ~,\nonumber
\end{align}
Note that the $x$-dependence vanishes in the zero frequency limit. Restoring $\hbar = h/2\pi=1$, the conductance reads\cite{apel_rice_82,kane_fisher_92}
\begin{align}
G&=\frac{2e^2}{h}K_{\rm{eff}} =\frac{e^2}{h}\widetilde{K}_{\rm{eff}} \label{eq:cond_inf}\\ &=\frac{e^2}{h} \frac{2K_c \sqrt{v_c v_s}\delta^2}{\sqrt{(K_c K_s v_c+v_s \delta^2)(v_c \delta^2+K_c K_s v_s)}}~.\nonumber
\end{align}
We thus find that the conductance of an infinite, homogeneous wire depends not only on the interaction parameters $K_c$ and $K_s$, but also on the ratio of the charge and spin velocities, as well as on $\delta = \gamma_s/\gamma_c$. The ratio of charge and spin velocities enters the conductance because the gap couples the spin and charge degrees of freedom in the wire. If these do not propagate at the same speed, a pure charge excitation (such as the electric current) cannot propagate as easily as if the velocities were the same. This is similar to the suppression of the inter band Ruderman-Kittel-Kasuya-Yosida (RKKY) exchange in multi subband wires\cite{meng_13} or the Coulomb drag between two wires.\cite{drag_literature1,drag_literature2,drag_literature3} The ratio of $\gamma_s$ and $\gamma_c$, on the other hand, encodes how much spin and charge is being gapped out by the sine-Gordon potential. Quite naturally, the conductance becomes smaller the more charge is gapped. As a crosscheck, we find that the conductance is given by $2e^2/h(1+\delta^{-2})$ for a non-interacting wire ($K_s=K_c=1$) with $v_c=v_s$, in agreement with the result presented in a recent publication on fractional helical wires.\cite{oreg_fractional_13} We will see in Sec.~\ref{sec:FLL} how this results gets modified in the presence of Fermi liquid leads. 

\subsubsection{Drude peak}
Another interesting quantity is the conductivity, $\sigma(\omega)$, which is accessible in optical experiments and does therefore not require transport through leads. This allows to deduce $\sigma(\omega) = 2e^2\,(\omega+i0^+)\,G_{cc}^{\rm R}(k=0,\omega)/\pi^2$ from the action describing the infinite system. As usual, the optical conductivity shows a Drude peak at zero frequency, $\sigma(\omega) = D\,\delta(\omega) + i\,\tilde{\sigma}(\omega)$ (with $\tilde{\sigma}(0) = 0$). We obtain its weight $D$ as

\begin{align}
D=\frac{2 e^2}{\hbar}\,v_{\rm{eff}}K_{\rm{eff}}=\frac{2 e^2}{\hbar}\,\frac{K_c v_c v_s\delta^{2}}{K_c K_s v_c  + v_s\delta^{2}}~.
\end{align}
Like the conductance, the weight of the Drude peak depends on the interactions inside the wire, on the fraction of the charge mode that is being gapped by the sine-Gordon potential, and on the ratio of spin and charge velocities. In weakly interacting quantum wires, where $v_s \approx v_F$, $v_c \approx v_F/K_c$ and $K_s\approx 1$ (with $v_F$ being the Fermi velocity), the weight of the Drude peak assumes the non-interacting form $D=(e^2/\hbar)\,v_F\,2/(1+\delta^{-2})$. For $|\delta| = 1$ and $v_c=v_s$, our results furthermore recover the low frequency limit of reference [\onlinecite{schuricht_12}] in which the optical conductivity of a quantum wire has been analyzed for finite $\omega$ and $\Delta$ upon neglecting the coupling\cite{braunecker_prb_09} between the gapped and the gapless sector of the theory. This coupling $\sim v_c-v_s$ in turn modifies the weight of the Drude peak $D$ as compared to reference [\onlinecite{schuricht_12}].

\subsubsection{Low energy density of states}
The density of states at low energies can be obtained from the electronic propagators of the interacting helical wire,

\begin{align}
\rho(\omega) =- \frac{1}{\pi}\,\mathfrak{Im}\left\{\sum_{\sigma} G^{\rm R}_{\sigma}(x\to0,\omega)\right\}~,\label{eq:dos}
\end{align}
where $G^{\rm R}_{\sigma}(x\to0,\omega)$ is the retarded propagator of an electron with spin $\sigma = \uparrow,\downarrow$. The frequency dependence of $\rho(\omega)$ can for instance be measured in tunneling experiments. In a partially gapped quantum wire, the low frequency density of states stems from the gapless sector. For a non-fractional helical wire, and again neglecting the coupling between the gapped and the gapless sector, the form of $\rho(\omega)$ has been discussed in references [\onlinecite{schuricht_12,bena_12}].

For sufficiently long wires at frequencies $v_F/L\ll|\omega| \ll \text{min}\{\Delta\sqrt{v_c},\Delta \sqrt{v_s}\}$, one can obtain the density of states from the Eq.~\eqref{eq:general_action} as sketched in appendix \ref{append:dos}. In the infinite gap limit, we find that the low-energy electronic density of states vanishes unless $|\delta| =1$. For a finite gap, the density of states is suppressed for frequencies smaller than the gap, but recovers the gapless value at frequencies larger than the gap.\cite{starykh,voit_98} The vanishing of $\rho(\omega)$ for $|\delta|\neq1$ can be understood as a direct consequence of fractionalization in the wire. For $|\delta|<1$, the sine-Gordon potential gaps out more charge than spin, and vice-versa for $|\delta| >1$. Only for $|\delta|  = 1$, equally much charge and spin are gapped. Since we started from a spinful Luttinger liquid with two degenerate sets of states (spin up and spin down), gapping out half of them results in exactly one gapless electronic mode in the wire. For $|\delta| \neq 1$, however, there is either too little gapless charge or too little gapless spin to form a full electronic state, and the density of states for electrons vanishes consequently. This result can also be related to the uncertainty principle associated with the ordered combination of fields $\phi_+ = \phi_c + \delta \theta_s$. Defining its canonically conjugate field as $\theta_+ = (\theta_c+\delta^{-1}\phi_s)/2$, and the remaining two fields as $\phi_-  = \phi_c - \delta \theta_s$ and $\theta_- = (\theta_c-\delta^{-1}\phi_s)/2$, we find that  the electronic Green's function $G^{\rm R}_{\sigma}(x\to0,\omega)$ involves exponentials of $r\phi_c+r\sigma\phi_s-\theta_c-\sigma\theta_s = \left[(\delta r -\sigma)\phi_+ + (\delta r +\sigma)\phi_-\right]/(2\delta)+(\delta r \sigma -1)\theta_+ -(\delta r \sigma + 1)\theta_-$, where $r = R,L\equiv\pm1$ labels contributions from right and left moving electrons, and $\sigma =\uparrow,\downarrow\equiv\pm1$ denotes the spin, see appendix \ref{append:dos}. For $\delta \neq 1/(r\sigma)$, and thus especially for $|\delta| \neq 1$, the electronic Green's function contains the field $\theta_+$. Being conjugate to $\phi_+$, the field $\theta_+$ strongly fluctuates, such that an expectation value of an exponential involving this field vanishes. We therefore find

\begin{align}
\rho(\omega) \sim \begin{cases} |\omega|^\alpha&,~|\delta| = 1\\0&,~|\delta|\neq1~,\end{cases}
\end{align}
where 
\begin{align}
\alpha = \frac{1}{4K_c} \frac{1+K_c^2 \left(K_s^2+4\right)+K_c K_s \left(\frac{v_s}{v_c}+\frac{v_c}{v_s}\right)}{\sqrt{\frac{v_s}{v_c}+K_c K_s}\sqrt{\frac{v_c}{v_s}+K_c K_s}}-1~.\label{eq:dos_exponent}
\end{align}
In the limit $v_s=v_c$, the density of states exponent reduces to
\begin{align}
\alpha_0 = \frac{\left[1+K_c \left(K_s-2 \right)\right]^2}{4K_c \left(1+K_c K_s \right)}~,
\end{align}
in agreement with references [\onlinecite{schuricht_12, bena_12}]. For $v_c \neq v_s$, however, we obtain a modified exponent for the density of states. This discrepancy can be traced back to the neglect of terms $\sim v_c-v_s$ connecting the gapped and the gapless sector in the action considered in these works.\cite{dos_remark} With the experimentally realistic values $K_c\approx 0.65$, $K_s\approx 1$, and $v_i = v_F/K_i$,\cite{yompol_09} the resulting discrepancy is as large as $(\alpha-\alpha_0)/\alpha_0 \approx 0.19$.

\section{Conductance of a finite size wire connected to Fermi liquid leads}\label{sec:FLL}
\begin{figure}
\includegraphics[width=0.3\textwidth]{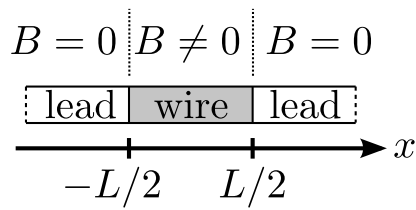}
\caption{Helical wire attached to leads from the left and right side at positions $x=\pm\frac{L}{2}$.}
\label{fig:wire}
\end{figure}

When measuring charge transport through a fractional helical Luttinger liquid, the unavoidable presence of Fermi liquid leads is expected to importantly modify the conductance.\cite{conductance_ll_leads1,conductance_ll_leads2,conductance_ll_leads3} We therefore now turn to an inhomogenous Luttinger liquid model, in which a fractional helical quantum wire of length $L$, located between $x=-L/2$ and $x=L/2$, is sandwiched between two Fermi liquid leads. This situation is depicted in Fig.~\ref{fig:wire}. We now show that the conductance of this system is independent of the Luttinger parameters $K_c$ and $K_s$ in the wire despite the entanglement of spin and charge degrees of freedom. The starting point is again the (matrix) Green's function $\mathcal{G}$, which is determined from

\begin{align}\label{eq:matrixeq}
\left( \begin{array}{cc}\hat{\mathcal{D}}_c(x,{\tau})& \delta \Delta^2(x)  \\ \delta \Delta^2(x)  & \hat{\mathcal{D}}_s(x,{\tau})\end{array}\right) \mathcal{G}(x,x') =\pi\,\delta(x-x')\,  \mathds{1} \;.
\end{align}
This differential equation is valid for all $x$ and $x'$. Inside the leads, the velocities and Luttinger parameters take the values $v_c^L$, $v_s^L$, $K_c^L$, $K_s^L$, while $B^L = 0$. For a Fermi liquid, all velocities equal the Fermi velocity, $v_c^L = v_s^L = v_F$, and $K_c^L = K_s^L =1$. For generality we stick to general lead parameters. In the wire, the parameters take the same values as before ($v_c$, $v_s$, $K_c$, $K_s$, and $B\neq 0$). Being interested in the conductance within the wire, we fix $x'\in[-L/2,L/2]$. The propagators can now be calculated along the lines of references [\onlinecite{conductance_ll_leads1,conductance_ll_leads3}] by solving the homogeneous differential equation \eqref{eq:matrixeq} away from the interfaces at $x=\pm L/2$ and $x=x'$, where discontinuities in the derivatives of the propagators occur (see below). The solutions are then matched at the interfaces. Inside the leads, each propagator satisfies a decoupled differential 
equation. The general solutions read

\begin{align}
G_{cc}(x,x',\omega_n) = a_{cc} \, e^{|\omega_n| x/v_c^L} + b_{cc} \, e^{-|\omega_n| x/v_c^L}~,\\
G_{cs}(x,x',\omega_n) = a_{cs} \, e^{|\omega_n| x/v_s^L} + b_{cs} \, e^{-|\omega_n| x/v_s^L}~,
\end{align}
and similar expressions for $G_{sc}$ and $G_{ss}$. The propagators have to vanish at infinity, which imposes $a_{ij} = 0$ (right lead) or $b_{ij} = 0$ (left lead). Inside the wire, the differential equations are pairwise coupled. In order to find the charge propagator for $x \neq x'$, we have to solve the differential equations

\begin{align}
&\hat{\mathcal{D}}_c(x,{\omega_n})\,G_{cc}(x,x',\omega_n) +\delta \Delta^2 \,G_{cs}(x,x',\omega_n) = 0~,\label{eq:diff1}\\
&\delta \Delta^2 \,G_{cc}(x,x',\omega_n) + \hat{\mathcal{D}}_s(x,{\omega_n})\,G_{cs}(x,x',\omega_n)= 0~\label{eq:diff2},
\end{align}
where $\hat{\mathcal{D}}_{c,s}(x,{\omega_n})$ are the Fourier transforms of the differential operators given in Eq.~\eqref{eq:diff_op}. One convenient way of solving these equations is to rewrite the system of two coupled partial differential equations of second order as a system of four coupled differential equations of first order by introducing the first order derivatives of the propagators, and solving for the eigenvalues and eigenvectors of the matrix associated with this system of coupled equations. One then finds that the propagators have the general solution

\begin{align}
G_{cc}(x,x',\omega_n) = \sum_{i=1}^4\, c_{i}\, \epsilon_{i,cc}\, e^{q_ix}~,\\
G_{cs}(x,x',\omega_n) = \sum_{i=1}^4\, c_{i}\, \epsilon_{i,cs}\, e^{q_ix}~,
\end{align}
where $c_i$ are four the constants of integration, $\epsilon_{i,cc}$ and $\epsilon_{i,cs}$ are components of the eigenvectors of the matrix associated with the system of coupled equations, while $q_i$ are the corresponding eigenvalues. The dependence on $x'$ is hidden in $c_{i}\, \epsilon_{i,cc}$ and $c_{i}\, \epsilon_{i,cs}$ (note that the differential operators in Eqs.~\eqref{eq:diff1} and \eqref{eq:diff2} only depend on $x$). To lowest order in $\omega_n/(\Delta\sqrt{v_F})$, we find that

\begin{align}
\frac{\epsilon_{1,cs}}{\epsilon_{1,cc}} &= \frac{\epsilon_{2,cs}}{\epsilon_{2,cc}} = -\frac{1}{\delta} + \mathcal{O}\left(\frac{\omega_n^2}{\Delta^2 v_F}\right)~,\\
\frac{\epsilon_{3,cs}}{\epsilon_{3,cc}} &= \frac{\epsilon_{4,cs}}{\epsilon_{4,cc}} = \frac{v_c \delta}{K_c  K_s  v_s }+  \mathcal{O}\left(\frac{\omega_n^2}{\Delta^2v_F}\right)~,
\end{align}
while the eigenvalues are given by
\begin{align}
q_1 &=-q_2 \\
&= |\omega_n|\,\sqrt{\frac{v_s \delta^2+K_c  K_s  v_c }{v_c  v_s (v_c \delta^2 +K_c  K_s  v_s )}}\left(1+ \mathcal{O}\left(\frac{\omega_n^2}{\Delta^2v_F}\right)\right)~,\nonumber\\
q_3 &= -q_4 =\Delta\,\sqrt{\frac{K_c }{v_c }+\frac{\delta^2}{K_s  v_s }}\left(1+ \mathcal{O}\left(\frac{\omega_n^2}{\Delta^2v_F}\right)\right)~.
\end{align}

\subsection{Matching the propagators at the interfaces} 
The matching of the propagators at the the interfaces at $x = \pm L/2$ and $x = x'$ finally provides the means to determine all constants of integration. At each interface, all propagators have to be continuous. At $x=\pm L/2$, the derivative of the propagators jump in such a way that the quantities $(v_c(x)/K_c(x)) \,\partial_x G_{cc}(x,x',\omega_n)$ and $v_s(x) K_s(x) \,\partial_x G_{cs}(x,x',\omega_n)$ remain continuous. At $x = x'$, the derivative of the mixed charge-spin propagator $\partial_x G_{cs}(x,x',\omega_n)$ is continuous, while the one of the charge propagator satisfies $\partial_x G_{cc}(x'+\epsilon,x',\omega_n) = \partial_x G_{cc}(x'-\epsilon,x',\omega_n) - \pi K_c(x')/v_c(x')$ for $\epsilon \to 0^+$. These conditions follow from integrating equation \eqref{eq:matrixeq} over the interfaces. For the conductance, which is position independent, a further simplification arises when one chooses $x'=0$, in which case the propagators have to be symmetric under inversion of $x$. This reduces the number of unknowns from 12 to 6. One may now solve for the unknowns and finds, again to leading order in $\omega_n/(\Delta\sqrt{v_F})$ and for $x\in[-L/2,0]$, that

\begin{align}
c_1\epsilon_{1,cc} &=\frac{\pi}{4|\omega_n|}\Bigl(\frac{ K_c^L\delta^{2}}{\delta^{2}+ K_c^L  K_s^L }\\
&+ \frac{K_c  \sqrt{v_c  v_s }\delta^{2}}{\sqrt{v_c \delta^{2}+K_c  K_s  v_s }\sqrt{v_s \delta^{2}+K_c  K_s  v_s}}\Bigr)+ \mathcal{O}\left(\omega_n^0\right) ~,\nonumber\\
c_2\,\epsilon_{2,cc} &= \frac{\pi}{4|\omega_n|}\Bigl(\frac{ K_c^L\delta^{2}}{1+ K_c^L  K_s^L}\\
&- \frac{K_c  \sqrt{v_c  v_s }\delta^{2}}{\sqrt{v_c \delta^{2}+K_c  K_s  v_s }\sqrt{v_s \delta^{2}+K_c  K_s  v_c }}\Bigr)+ \mathcal{O}\left(\omega_n^0\right)~,\nonumber\\
c_3\,\epsilon_{3,cc}&~,~~c_4\,\epsilon_{4,cc} = \mathcal{O}\left(\omega_n^0\right)~.% \mathcal{O}\left(\frac{\omega_n^2}{\Delta^2}\right)~.
\end{align}
Using these results, we can finally evaluate the conductance according to Eq.~\eqref{eq:cond}, which reads

\begin{align}
G = \frac{2e^2}{h}\,\frac{ K_c^L}{1+ K_c^L  K_s^L \,\delta^{-2}}~.\label{eq:cond2}
\end{align}
For a regular helical Luttinger liquid, characterized by $|\delta|=1$, we thus recover a quantized conductance of $G=e^2/h$ (this value however only holds at zero temperature\cite{schmidt_finite_t}), while $\delta^{-1} = 2n+1$ recovers the result of Ref.~[\onlinecite{oreg_fractional_13}], where the non-integer value of the conductance has been found in a scattering matrix approach that did not explicitly address electron-electron interactions inside the wire. It is interesting to note that the form of the conductance in Eq.~\eqref{eq:cond2} does not follow from the conductance of the infinite wire, given in Eq.~\eqref{eq:cond_inf}, upon replacing all wire parameters by the corresponding lead parameters, such as $K_c  \to  K_c^L$. Instead, the lead velocities $v_{c}^L$ and $v_{s}^L$ do not appear in Eq.~\eqref{eq:cond2}. This is due to a cancellation of the lead velocities in the interface condition at $\pm L/2$ with an inverse factor of lead velocity stemming form the derivative of the propagator. The lead velocities therefore only appear in the exponents of the propagators as length dependent combinations $|\omega_n| L / v_{c}^L$ and $|\omega_n| L / v_{s}^L$, which cannot contribute to the conductance. Put in a more physical language, the velocities $v_c^L$ and $v_s^L$ would only enter the expression of the conductance if a non-vanishing gap existed inside the leads, in which case the same discussion as for the infinite wire would apply. In the present setup, however, spin and charge are decoupled inside the leads, so that a velocity mismatch between the two is irrelevant for pure charge transport.

\section{Finite frequency conductivity}\label{sec:finite_freq}
In the last sections, we have employed an expansion of the sine-Gordon potential to second order, which brings the Hamiltonian to a quadratic form. The latter in turn allowed the calculation of observables for energies much smaller than the helical gap. This treatment, however, breaks down for physical properties associated with energies of the order of the gap. At energies much larger than the gap, on the other hand, the sine-Gordon potential becomes completely irrelevant, and so does the error due to its quadratic expansion. The expansion should thus be accurate both at energies much smaller and much larger than the gap.

\subsection{Frequencies much smaller than the gap}\label{subsec:smaller}
At frequencies much smaller than the gap, but still larger than the finite size frequency $\omega_L = v_F/L$, the discussion of Sec.~\ref{sec:inifite_wire} implies that a fractional helical Luttinger liquid should exhibit the charge transport properties of a spinless Luttinger liquid with Luttinger parameter $\widetilde{K}_{\rm eff} = 2 K_{\rm eff}$ and effective velocity $v_{\rm eff}$ as defined in Eq.~\eqref{eq:effparam}. In this picture, the reduced value of the conductance given in Eq.~\eqref{eq:cond2} can be understood as coupling this spinless Luttinger liquid to spinless and interacting Luttinger liquid leads of velocity $v_F$ and Luttinger parameter $K_{{\rm eff}, L} = 2K_{c,L}/(1+K_{c,L}K_{s,L}\delta^{-2})$. 

\begin{figure}
(a)~\includegraphics[width=0.49\textwidth]{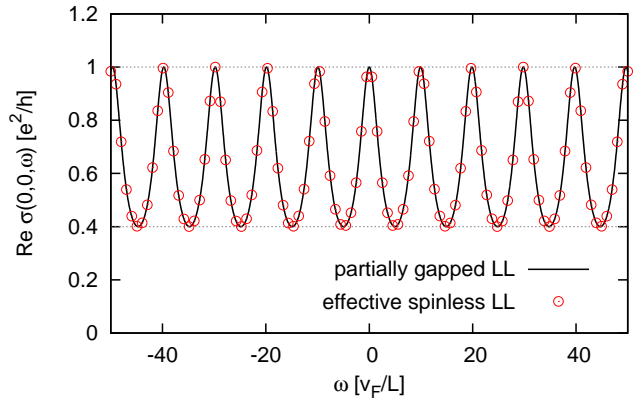}\\
(b)~\includegraphics[width=0.49\textwidth]{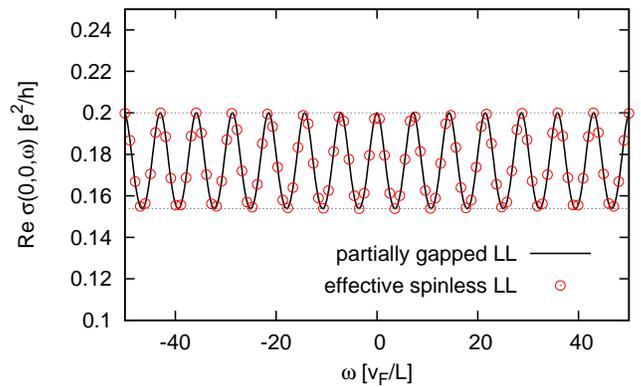}
\caption{Real part of the conductivity $\mathfrak{Re}\left\{\sigma(0,0,\omega)\right\}$ as defined in Eq.~\eqref{eq:condj} at frequencies much smaller than the gap. The solid line is the numerically evaluated conductivity for $\Delta = 10000\,\sqrt{v_F}/L$, $K_c = 0.5$, $K_s = 1$, and $v_{i} = v_F/K_i$. Panel (a) shows data for $\delta = 1$, panel (b) corresponds to $\delta = 1/3$. The superimposed circles have been calculated from the known analytical expression\cite{conductance_ll_leads1,conductance_ll_leads2,conductance_ll_leads3} of $\sigma(0,0,\omega)$ for a spinless Luttinger liquid (LL) with effective Luttinger parameter $\widetilde{K}_{\rm eff}$ and effective velocity $v_{\rm eff}$ coupled to spinless Luttinger liquid leads of velocity $v_{F}$ and Luttinger parameter $K_{{\rm eff},L}$,  while the horizontal dashed lines indicate the values between which this latter analytical expression oscillates, see main text.}
\label{fig:gapped}
\end{figure}

To demonstrate this behavior, we analyze the non-local conductivity $\sigma(x,x',\omega)$ at frequency $\omega$ between the points $x$ and $x'$ along the wire. The latter is defined by the relation between the electric field $E(x',\omega)$ and the current $j(x,\omega)$,

\begin{align}
j(x,\omega) = \int dx'\,\sigma(x,x',\omega)\,E(x',\omega)~.\label{eq:condj}
\end{align}
In linear response, it is given by

\begin{align}
\sigma(x,x',\omega) =\left.\frac{2e^2}{\pi^2}\,\omega_n\,\langle\phi_c(x,\omega_n)\phi_c(x',-\omega_n)\rangle\right|_{i\omega_n \to \omega+i0^+} ~,
\end{align}
and can be evaluated similarly to the conductance, which is in fact given by the limit $G = \lim_{\omega\to0} \sigma(x,x',\omega)$ (in this limit, the non-local conductivity becomes independent of $x$ and $x'$). Figure \ref{fig:gapped} contrasts the frequency dependence of the non-local conductivity $\sigma(0,0,\omega)$, calculated by numerically matching the charge propagator at the interfaces $x=\pm L/2$ and $x=x'$ as described in Sec.~\ref{sec:FLL}, with the well-known conductivity for a spinless Luttinger liquid connected to interacting Luttinger liquid leads, which has been established in references [\onlinecite{conductance_ll_leads1,conductance_ll_leads2,conductance_ll_leads3}]. From this latter analytical formula, one can directly infer that the non-local conductivity oscillates between $\sigma(0,0,\omega) = K_{{\rm eff},L}\,e^2/h$ at $\omega_{\rm max} = 2n\pi\, u_{\rm eff}/L$ and $\sigma(0,0,\omega) = (\widetilde{K}_{\rm eff}^2/K_{{\rm eff},L})\,e^2/h$ at $\omega_{\rm min} = (2n+1)\pi\, u_{\rm eff}/L$ (with $n\in\mathbb{Z}$). Figure \ref{fig:gapped} clearly demonstrates that the low-energy charge transport properties of a fractional helical Luttinger liquid can indeed be understood as the ones of a spinless Luttinger liquid of velocity $v_{\rm eff}$ and Luttinger parameter $\widetilde{K}_{\rm eff}$, coupled to interacting Luttinger liquid leads of velocity $v_F$ and Luttinger parameter $K_{{\rm eff}, L}$.

\subsection{Frequencies much larger than the gap} \label{subsec:larger}
\begin{figure}
\includegraphics[width=0.49\textwidth]{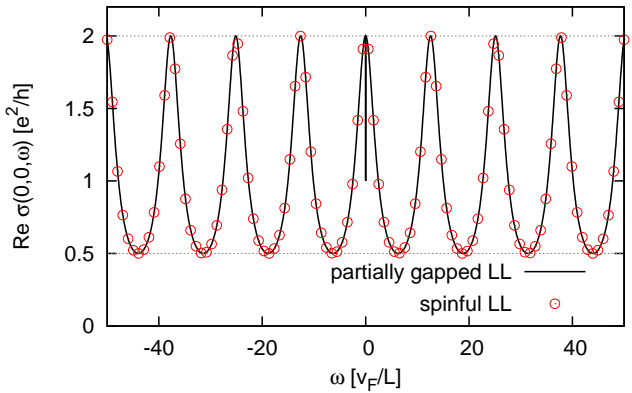}
\caption{Real part of the conductivity $\mathfrak{Re}\left\{\sigma(0,0,\omega)\right\}$ at frequencies much larger than the gap. The solid line corresponds to $\Delta = 0.01\,\sqrt{v_F}/L$, $K_c = 0.5$, $K_s = 1$, $v_{i} = v_F/K_i$, and $\delta = 1$. The circles stem from the known expression\cite{conductance_ll_leads1,conductance_ll_leads2,conductance_ll_leads3} of $\sigma(0,0,\omega)$ for a spinful Luttinger liquid (LL) with the same parameters coupled to (here non-interacting) Luttinger liquid leads. Like in Fig.~\ref{fig:gapped}, the horizontal dashed lines indicate the values between which this latter expression oscillates. For $\omega\to 0$, the conductivity reaches the gapped value $1\,e^2/h$ in a sharp drop. Different values of $\delta$ lead to the same curve, up to a modified zero frequency dip that recovers the conductance given in Eq.~\eqref{eq:cond2}.}
\label{fig:gapless}
\end{figure}

At frequencies much larger than the gap, the sine-Gordon term constitutes a subleading energy scale. The system should therefore behave as a regular spinful Luttinger liquid coupled to Fermi liquid leads. Given that we work close to equilibrium, this can best be checked by reducing the gap whilst keeping all other quantities fixed. Figure \ref{fig:gapless} depicts the real part of the numerically evaluated finite frequency conductivity $\mathfrak{Re}\left\{\sigma(0,0,\omega)\right\}$ for a very small gap, in comparison to the well-known analytic result\cite{conductance_ll_leads1,conductance_ll_leads2,conductance_ll_leads3} for a spinful quantum wire, and no difference is visible apart from a sharp drop of the conductivity for $\omega\to0$. With this drop, the conductivity recovers the conductance given in Eq.~\eqref{eq:cond2} at zero frequency.

\subsection{Crossover regime}\label{subsec:inter}
\begin{figure}
\includegraphics[width=0.49\textwidth]{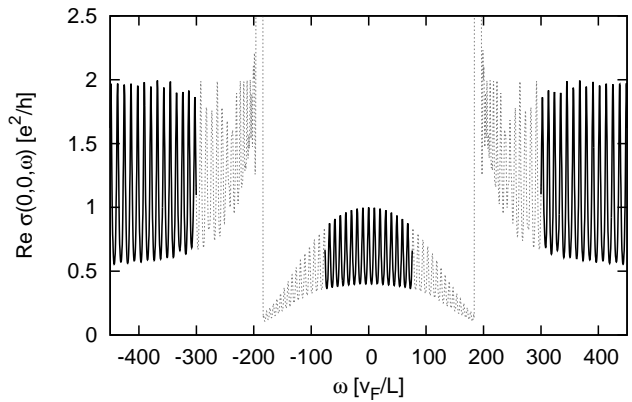}
\caption{Real part of the conductivity $\mathfrak{Re}\left\{\sigma(0,0,\omega)\right\}$, evaluated numerically for $\Delta = 130\,\sqrt{v_F}/L$, $K_c = 0.5$, $K_s = 1$, $v_{i} = v_F/K_i$, and $\delta = 1$ (this value of $\Delta$ corresponds to a gap of $\omega_{\rm gap} \approx 184\,v_F/L$). At small frequencies, the behavior of a spinless Luttinger liquid is recovered, while the conductivity approaches the expression of a spinful and gapless Luttinger liquid at large frequencies. For frequencies close to the gap, not only the expansion of the sine-Gordon potential is unjustified, but the conductivity also shows some unphysical, noisy features (sections drawn with a gray dashed line).}\label{fig:crossover}
\end{figure}

For completeness, we also show the real part of the conductivity $\mathfrak{Re}\left\{\sigma(0,0,\omega)\right\}$ at frequencies of the order of the gap. Figure \ref{fig:crossover} illustrates that our approach yields an expression of $\sigma(0,0,\omega)$ that recovers and interconnects the limits of small gap and large gap. Given that the gap energy is much larger than the finite size energy $\omega_L = v_F/L$, one can check from the action describing the infinite wire system, see Eq.~\eqref{eq:lowenergy_action}, that the gap is given by

\begin{align}
\omega_{\rm gap} = \Delta \sqrt{v_c K_c + \frac{v_s}{K_s}}~,
\end{align}
which simplifies to $\omega_L = \sqrt{2 v_F}\,\Delta$ for $v_i = v_F/K_i$ and $K_s=1$. We would like to stress, however, that the expansion of the cosine to second order used to calculate $\sigma(0,0,\omega)$ renders the values obtained for $\sigma(x,x,\omega\approx\omega_{\rm gap})$ meaningless. For frequencies very close to the gap, the conductivity furthermore shows noisy features, including an unphysical overshooting to values larger than $2\,e^2/h$, which might be due to its numerical evaluation.

\section{Conclusions}\label{sec:sum}
In this work, we have analyzed the conductance and finite frequency conductivity, the Drude peak and the density of states of a partially gapped quantum wire. Similar to a gapless wire, the presence of Fermi liquid leads strongly modifies the conductance, and brings it from the non-universal value given in Eq.~\eqref{eq:cond_inf}, which depends on the effective velocities and Luttinger liquid parameters inside the wire, to the form of Eq.~\eqref{eq:cond2}. In particular, the conductance of a (fractional) helical wire connected to leads does not depend on the effective velocities and Luttinger liquid parameters inside the interacting wire. Its expression does, however, not follow from the one of an infinite, homogeneous wire upon replacing all wire parameters by lead parameters (such as $K_c \to  K_c^L$). This is different from a gapless quantum wire.\cite{conductance_ll_leads1,conductance_ll_leads2,conductance_ll_leads3} In the latter case, the interacting wire can be smoothly deformed into a gapless Fermi liquid. Such a smooth deformation cannot be found between a partially gapped and a fully gapless state. Furthermore, the conductance of a partially gapped quantum wire also depends on the precise form of the sine-Gordon potential that drives the opening of the partial gap, as has already been established in Ref.~[\onlinecite{oreg_fractional_13}]. As a consequence, the number of gapless modes is not sufficient to evaluate the conductance once spin and charge degrees of freedom get mixed. For a sine-Gordon potential of the form $\cos\left(\sqrt{2}(\gamma_c\,\phi_c+\gamma_s\,\theta_s)\right)$, the conductance reads $G=(e^2/h)\,2/(1+(\gamma_c/\gamma_s)^2)$. By symmetry of $\theta_s\leftrightarrow\phi_s$ along with $K_s\leftrightarrow 1/K_s$, we furthermore conclude that a sine-Gordon potential of the type $\cos\left(\sqrt{2}(\gamma_c\,\phi_c+\gamma_s'\,\phi_s)\right)$ would yield a similar conductance of $G=(e^2/h)\,2/(1+(\gamma_c/\gamma_s')^2)$. The conductance is thus in general not quantized in integer multiples of $e^2/h$, which can be understood as a signature of charge fractionalization.\cite{oreg_fractional_13} The general form of the conductance given in in Eq.~\eqref{eq:cond2} also reproduces the conductance $G=0$ of a Mott insulator (where $\gamma_s = 0$), the value $G=2\,e^2/h$ of a spin-density wave state (where $\gamma_c =0$), and the conductance $G=1\,e^2/h$ of a spin polarized system (the latter case would correspond to a sine-Gordon potential of the form $\cos(\sqrt{2}(\gamma_c \phi_c \pm \gamma_s' \phi_s))$ with $\gamma_c =\gamma_s'$). In addition to the conductance, we have also calculated the Drude peak of the optical conductivity of (quasi) helical and fractional helical wires. Finally analyzing the non-local conductivity of fractional helical Luttinger liquids at finite frequencies, we have furthermore shown that the low frequency charge transport properties of a fractional helical Luttinger liquid can be understood as the ones of a spinless Luttinger liquid coupled to interacting Luttinger liquid leads. This analogy, however, does not hold for all properties. The tunneling density of states is for instance only non-zero for regular helical Luttinger liquids without fractionalization. At frequencies larger than the gap, the wire finally behaves as a gapless, spinful Luttinger liquid coupled to Fermi liquid leads. The recovery of the gapless behavior is explicitly contained in our treatment.

 \acknowledgements
We would like to thank B. Braunecker, J. Klinovaja, V. Meden, A. Rosch, A. Saha, T. Schmidt, and P. Simon for helpful discussions. This work has been supported by SNF, NCCR Nano, and NCCR QSIT (TM and DL), as well as by the German Research Foundation (DFG) through the Emmy-Noether program under FR 2627/3-1 (LF) and SCHU 2333/2-1 (DS).

\appendix
\section{Density of states}\label{append:dos}
We start from the general expression of the frequency-dependent density of states $\rho(\omega)$ given in Eq.~\eqref{eq:dos}. We linearize the spectrum of the quantum wire around its two Fermi points at momentum $\pm k_F$, and decompose the operator $c_{\sigma}(x)$, which annihilates an electron of spin $\sigma$ at position $x$, into its right and left moving parts, $c_{\sigma}(x) = e^{i xk_F} R_{\sigma}(x) + e^{-i xk_F} L_{\sigma}(x)$.  The latter can be bosonized as $r_{\sigma}(x) =(U_{r\sigma}/\sqrt{2\pi a})\,e^{-i(r\phi_{\sigma}(x)-\theta_{\sigma}(x))}$, where $r=R,L\equiv\pm1$, while the corresponding Klein factors are denoted as $U_{r\sigma}$. \cite{giamarchi_book} We also introduce the spin and charge fields as $\phi_{\hspace*{-2.5pt}\begin{array}{c}\\[-19.5pt] {}_{c}\\[-7pt] {}_{s}\end{array}}(z) = (\phi_{\uparrow}\pm\phi_{\downarrow})/\sqrt{2}$ and $\theta_{\hspace*{-2.5pt}\begin{array}{c}\\[-19.5pt] {}_{c}\\[-7pt] {}_{s}\end{array}}(z) = (\theta_{\uparrow}\pm\theta_{\downarrow})/\sqrt{2}$. With these definitions, the imaginary time Green's function of an electron of spin $\sigma$ reads

\begin{align}
&G_{\sigma}(x,\tau) = -\langle T_{\tau}\,c_{\sigma}^\dagger(x,\tau)c^\pdag_{\sigma}(0,0)\rangle\\
%&=\sum_{r}\frac{-1}{2\pi\alpha}\,\langle T_\tau \,e^{i(r\phi_\sigma(x,\tau)-\theta_\sigma(x,\tau))-i(r\phi_\sigma(0,0)-\theta_\sigma(0,0))}  \rangle\nonumber\\
&=\sum_{r}\frac{-1}{2\pi a}\,\langle T_\tau \,e^{i r (\phi_c(x,\tau)+\sigma \phi_s(x,\tau) - \phi_c(0,0) -\sigma \phi_s(0,0) )/\sqrt{2}}\nonumber\\
&\quad\quad\quad\quad\quad\times e^{-i (\theta_c(x,\tau)+\sigma \theta_s(x,\tau) - \theta_c(0,0) -\sigma \theta_s(0,0) )/\sqrt{2}}\rangle\nonumber~,
\end{align}
where we used $U_{r\sigma}^{\dagger} U_{r\sigma}^\pdag = 1$. Introducing

\begin{align}
\boldsymbol{q} =\frac{1}{\sqrt{2}}\, \begin{pmatrix}r\\-\sigma\\-1\\r\sigma\end{pmatrix}\quad,\quad\boldsymbol{\phi}(x',\tau') = \begin{pmatrix}\phi_c(x',\tau')\\ \theta_s(x',\tau')\\ \theta_c(x',\tau')\\ \phi_s(x',\tau') \end{pmatrix}~,
\end{align}
and defining

\begin{align}
\boldsymbol{q}(x',\tau') = \boldsymbol{q}\, \left[\delta(x'-x)\delta(\tau'-\tau)-\delta(x')\delta(\tau')\right]
\end{align}
we obtain

\begin{align}
&G_{\sigma}(x,\tau)=-\frac{1}{2\pi a}\\
&\times\sum_r\langle T_\tau\,e^{i\int dx'd\tau'\,[\boldsymbol{q}(x',\tau') \boldsymbol{\phi}^T(x',\tau') + \boldsymbol{\phi}(x',\tau') \boldsymbol{q}^T(x',\tau')]/2}\rangle~.
\nonumber
\end{align}
After Fourier transformation of the fields,
\begin{align}
\boldsymbol{\phi}(k,\omega_n) = \frac{1}{\sqrt{\beta L}}\int dx' \,d\tau'\,e^{i(\omega_n\tau'-kx')}\,\boldsymbol{\phi}(x',\tau')~,
\end{align}
where $\beta = 1/T$ is the inverse temperature and $L$ the length of the wire, and an analogous Fourier transformation for $\boldsymbol{q}(x',\tau')$, we obtain
\begin{align}
&G_{\sigma}(x,\tau)=-\frac{1}{2\pi a}\\
&\times\sum_r \langle e^{i\sum_{k,\omega_n}\,[\boldsymbol{q}_{-k,-\omega_n}^T\boldsymbol{\phi}_{k,\omega_n} + \boldsymbol{\phi}_{-k,-\omega_n}^T \boldsymbol{q}_{k,\omega_n}]/2}\rangle\nonumber~.
\end{align}
This expression corresponds to the field integral

\begin{align}
&G_{\sigma}(x,\tau)=-\frac{1}{2\pi a}\\
&\times\sum_r \int \frac{D(\boldsymbol{\phi}^T,\boldsymbol{\phi})}{Z}\,e^{-\frac{1}{2}\sum_{k,\omega_n}\,[\boldsymbol{\phi}^T\mathcal{G}^{-1}\boldsymbol{\phi}-i\boldsymbol{q}^T\boldsymbol{\phi}-i\boldsymbol{\phi}^T\boldsymbol{q}]_{k,\omega_n}}~,\nonumber
\end{align}
where $Z$ is the partition function, while the inverse matrix Green's function reads

\begin{align}
\mathcal{G}^{-1}_{k,\omega_n} &=\frac{1}{\pi}\begin{pmatrix}\frac{k^2 v_c}{K_c}+\Delta^2& \delta \Delta^2&i k \omega_n&0\\\delta\Delta^2&k^2 v_s K_s+\delta^2\Delta^2&0&i k \omega_n\\i k \omega_n&0&k^2 v_c K_c&0\\0&i k \omega_n&0&\frac{k^2 v_s}{K_s}\end{pmatrix}~.
\end{align}
We can now integrate out $\boldsymbol{\phi}-i\mathcal{G}\boldsymbol{q}$ and obtain

\begin{align}
G_{\sigma}(x,\tau)&=\frac{-1}{2\pi a} \sum_r e^{-\frac{1}{2}\sum_{k,\omega_n}\boldsymbol{q}_{-k,-\omega_n}^T\mathcal{G}_{k,\omega_n}\boldsymbol{q}_{k,\omega_n}}\\
&=\frac{-1}{2\pi a}\sum_r e^{-\sum_{k,\omega_n}\boldsymbol{q}^T\mathcal{G}_{k,\omega_n}\boldsymbol{q}\,[1-\cos(\omega_n\tau-kx)]}~.\nonumber
\end{align}
Being interested in the low frequency limit of the density of states, we can take the limit $\Delta\to\infty$. The exponent can then be written as a sum of three terms,

\begin{align}
&\frac{1}{\beta L}\sum_{k,\omega_n}\boldsymbol{q}^T\mathcal{G}_{k,\omega_n}\boldsymbol{q}\,[1-\cos(\omega_n\tau-kx)]\\
=&\frac{1}{\beta L}\sum_{k,\omega_n}\,(X_0+X_1+X_2)\,[1-\cos(\omega_n\tau-kx)]\nonumber~,
\end{align}
with $X_i \sim (\omega_n/k)^i$. The term $X_2$ is given by
\begin{align}
X_2 = \frac{\omega_n^2}{2k^2}\frac{\pi K_s (r\sigma\delta-1)^2}{k^2v_c v_s (K_c K_s v_s+v_c \delta^2)+\omega_n^2(K_c K_s v_c+v_s\delta^2)}~.
\end{align}
Summing this term over Matsubara frequencies and momenta yields a divergent contribution to the exponent,\cite{starykh,voit_98} which in turn suppresses the density of states (at finite but large $\Delta$, the Green's function $G(x,\omega_n)$ is suppressed as\cite{starykh,voit_98} $1/\Delta$). As a crosscheck, we note that Rashba nanowires subject to a magnetic field perpendicular to the direction set by the spin-orbit interaction correspond to $|\delta| = 1$.\cite{braunecker_prb_10} In these systems, the helical gap around zero momentum\cite{streda_03} leads to a reduction of the density of states by a factor of two as compared to gapless wires. This reduction is precisely due to the term $X_2$, which kills two out of four contributions to the density of states, namely the ones with $r = -\sigma = \pm \delta$. For frequencies $\omega$ smaller than the gap, we thus find that density of states of a general helical wire, given by the imaginary part of the analytic continuation of $G(x,\omega_n)$, is only non-zero if

\begin{align}
\delta = r \sigma~,
\end{align}
which in particular implies $|\delta|=1$. Put differently, an electron can only tunnel into the wire if there is a non-zero gapless density of states for a full electron, including both its full charge and its full spin. This is the case for $|\delta|=1$, while at $|\delta| \neq1$, there is either too little gapless spin or too little gapless charge to form a full electronic state. Only for $|\delta| = 1$, there are two combinations of $r$ and $\sigma$ with $\delta = r \sigma$ such that $X_2 = 0$. In this case, we furthermore find that

\begin{align}
X_1&=\frac{\omega_n}{k}\,\frac{2i\pi r(K_cK_sv_c+v_s)}{k^2 v_c v_s(v_c+K_cK_sv_s)+\omega_n^2 (v_s+K_cK_sv_c)}
\end{align}
results in a phase factor multiplying the density of states,\cite{giamarchi_book} while the most important contribution to $\rho(\omega)$ is given by

\begin{align}
X_0 &= \frac{2\pi\tilde{u}\widetilde{K}}{\tilde{u}^2k^2+\omega_n^2}~,
\end{align}
with
\begin{align}
\tilde{u} &=\sqrt{v_c v_s}\,\sqrt{\frac{v_c+K_cK_sv_s}{v_s+K_cK_sv_c}}~,\\
\widetilde{K} &= \frac{1}{4K_c} \frac{1+K_c^2 \left(K_s^2+4\right)+K_c K_s \left(\frac{v_s}{v_c}+\frac{v_c}{v_s}\right)}{\sqrt{\frac{v_s}{v_c}+K_c K_s}\sqrt{\frac{v_c}{v_s}+K_c K_s}}~.
\end{align}
The contribution of this term to the exponent is given by\cite{giamarchi_book}

\begin{align}
&\frac{1}{\beta L}\sum_{k,\omega_n}\,\frac{2\pi\tilde{u}\widetilde{K}}{\tilde{u}^2k^2+\omega_n^2}\,[1-\cos(\omega_n\tau-kx)]\\
&=\widetilde{K}\ln\left(\frac{\sqrt{x^2+(\tilde{u}|\tau|+a)^2}}{a}\right)~.
\end{align}
We thus find that

\begin{align}
G_\sigma(x,\tau) \sim \left(\frac{a}{\sqrt{x^2+(\tilde{u}|\tau|+a)^2}}\right)^{\widetilde{K}}~.
\end{align}
After analytic continuation\cite{giamarchi_book} and Fourier transformation, we finally obtain the density of states as

\begin{align}
\rho(\omega) = \sum_\sigma G^{\rm R}_{\sigma}(x\to 0,\omega) \sim |\omega|^{\widetilde{K}-1}~.
\end{align}

%%%%%%%%%%%%%%%%%%%%%%%


\begin{thebibliography}{99}
\bibitem{prb_remark}
In reference [\onlinecite{braunecker_prb_09}], the effective velocity and the effective Luttinger parameter are given in Eqs.~(61) and (62) (note that $\Gamma$ given in Eq.~(62) should be corrected to $4 \Gamma$), modulo a rescaling of the fields by a factor of $\sqrt{K_c/(K_cK_s+1)}$.
\bibitem{dos_remark}
We have checked that the calculation of reference [\onlinecite{bena_12}] yields the density of states exponent given in Eq.~\eqref{eq:dos_exponent} if one keeps track of the coupling between the gapped and the gapless sector. This can be done by replacing $K_-=1$ used in reference [\onlinecite{bena_12}] by $K_-^{\rm eff}$ as given in Eqs.~(61) and (62) of reference [\onlinecite{braunecker_prb_09}], modulo the correction $\Gamma\to4\Gamma$.\cite{prb_remark} The full result is thus independent of the choice of basis and other details of the calculation, as it should be.
\end{thebibliography}
\end{document}